\documentclass[iop]{emulateapj}

\slugcomment{}

\shorttitle{A possible FRB/GRB connection}
\shortauthors{Zhang}

\begin{document}

%% LaTeX will automatically break titles if they run longer than
%% one line. However, you may use \\ to force a line break if
%% you desire.

\title{A possible connection between Fast Radio Bursts and Gamma-Ray Bursts}
%FRB/GRB connection: towards a 
%multi-wavelength campaign to unveil the nature
%of Fast Radio Bursts}

%% Use \author, \affil, and the \and command to format
%% author and affiliation information.
%% Note that \email has replaced the old \authoremail command
%% from AASTeX v4.0. You can use \email to mark an email address
%% anywhere in the paper, not just in the front matter.
%% As in the title, use \\ to force line breaks.

\author{Bing Zhang}
%\altaffilmark{1,2,3}}
% and Ivan R. King\altaffilmark{1}}
%\affil{Kavli Institute of Astronomy and Astrophysics, Peking University, 
%Beijing 100871, China}
%\affil{Department of Astronomy, Peking University, Beijing 100871, China}
\affil{Department of Physics and Astronomy, University of Nevada Las Vegas, NV 89154, USA}

%% Notice that each of these authors has alternate affiliations, which
%% are identified by the \altaffilmark after each name.  Specify alternate
%% affiliation information with \altaffiltext, with one command per each
%% affiliation.

%\altaffiltext{1}{Kavli Institute of Astronomy and Astrophysics, 
%Peking University, Beijing 100871, China}
%\altaffiltext{2}{Department of Astronomy, Peking University, 
%Beijing 100871, China}
%\altaffiltext{3}{Department of Physics and Astronomy, University of 
%Nevada Las Vegas, NV 89154, USA}

%% Mark off your abstract in the ``abstract'' environment. In the manuscript
%% style, abstract will output a Received/Accepted line after the
%% title and affiliation information. No date will appear since the author
%% does not have this information. The dates will be filled in by the
%% editorial office after submission.

\begin{abstract}
The physical nature of Fast Radio Bursts (FRBs), a new type of cosmological
transients discovered recently, is not known.
It has been suggested that FRBs can be produced when a spinning supra-massive
neutron star loses centrifugal support and collapses to a black hole.
Here we suggest that such implosions can happen in 
supra-massive neutron stars shortly (hundreds to thousands of seconds) after 
their births, and an observational signature of such implosions 
may have been observed in the X-ray 
afterglows of some long and short gamma-ray bursts (GRBs). Within
this picture, a small fraction of FRBs would be physically connected to GRBs. 
We discuss possible multi-wavelength electromagnetic signals and gravitational 
wave signals that might be associated with FRBs, and propose an
observational campaign to unveil the physical nature of FRBs.
In particular, we strongly encourage a rapid radio follow-up observation of 
GRBs starting from 100 s after GRB triggers.
\end{abstract}

%% Keywords should appear after the \end{abstract} command. The uncommented
%% example has been keyed in ApJ style. See the instructions to authors
%% for the journal to which you are submitting your paper to determine
%% what keyword punctuation is appropriate.


%% From the front matter, we move on to the body of the paper.
%% In the first two sections, notice the use of the natbib \citep
%% and \citet commands to identify citations.  The citations are
%% tied to the reference list via symbolic KEYs. The KEY corresponds
%% to the KEY in the \bibitem in the reference list below. We have
%% chosen the first three characters of the first author's name plus
%% the last two numeral of the year of publication as our KEY for
%% each reference.

%% Authors who wish to have the most important objects in their paper
%% linked in the electronic edition to a data center may do so by tagging
%% their objects with \objectname{} or \object{}.  Each macro takes the
%% object name as its required argument. The optional, square-bracket 
%% argument should be used in cases where the data center identification
%% differs from what is to be printed in the paper.  The text appearing 
%% in curly braces is what will appear in print in the published paper. 
%% If the object name is recognized by the data centers, it will be linked
%% in the electronic edition to the object data available at the data centers  
%%
%% Note that for sources with brackets in their names, e.g. [WEG2004] 14h-090,
%% the brackets must be escaped with backslashes when used in the first
%% square-bracket argument, for instance, \object[\[WEG2004\] 14h-090]{90}).
%%  Otherwise, LaTeX will issue an error. 

\section{Introduction}

Recently, a new type of cosmological transients, dubbed Fast Radio Bursts (FRBs),
was discovered \citep{lorimer07,thornton13}. These radio bursts have a typical
duration of several milli-seconds, high Galactic latitudes, and anomalously 
high dispersion measure (DM) values corresponding to a cosmological 
redshift $z$ between 0.5 and 1 \citep{thornton13}. The inferred total 
energy release is $10^{38} - 10^{40}$ ergs, and the peak radio luminosity is 
$\sim 10^{43}~{\rm erg~s^{-1}}$. No detected electromagnetic counterpart was
claimed to be associated with FRBs.

The physical nature of FRBs is unknown. \cite{thornton13} discussed several 
possibilities, and suggested that the event rate of FRBs ($R_{\rm FRB}\sim 10^{-3} 
~{\rm gal}^{-1} {\rm yr}^{-1}$) is much higher than those of of gamma-ray bursts
(GRBs) and compact star mergers, but could be consistent with those of 
soft gamma-ray repeater giant flares or core-collapse supernovae.
Since the announcement of the discovery, several proposals
have been made to interpret FRBs, including delayed collapses of supra-massive
neutron stars to black holes \citep{falcke13},
special magnetar radio flares \citep{popov07},
mergers of double neutron stars \citep{totani13},
mergers of binary white dwarfs \citep{kashiyama13},
and flaring stars \citep{loeb13}.

\section{FRBs as implosions of new-born supra-massive neutron stars}

The milli-second duration $\tau$ points towards a small emission size 
for FRBs: $r_{_{\rm FRB}} \sim c \tau \sim 3\times 10^7~{\rm cm}~ (\tau/{\rm ms})$.
The source of emission has to be limited to very compact objects involving
neutron stars or black holes (white dwarfs may be marginally accommodated,
Kashiyama et al. 2013). At such a small size, the brightness temperature of 
radio emission is extremely high, so the radiation mechanism must be coherent
\citep{katz13}.

\cite{falcke13} made a good case that a supra-massive neutron
star collapsing into a black hole would be a likely source of FRBs. The supra-massive
neutron star is initially sustained centrifugally by rapid rotation. As
it gradually spins down, it would collapse into a black hole
when centrifugal support no longer holds gravity. When the magnetic
field ``hair'' is ejected as the event horizon swallows the neutron star,
a strong electromagnetic signal in the radio band (which they call 
a ``blitzar'') is released. This is an FRB. 

\cite{falcke13} suggested that such a delayed collapse would happen
several-thousand to million years after the birth of the supra-massive neutron
star. Here we propose that a small fraction of such implosions could also 
happen shortly (hundreds to thousands of seconds) after the birth of the neutron 
star, and a signature of such implosions may have been observed in the early
X-ray afterglow light curves of some GRBs.

GRBs may originate from two types of progenitor: collapse of a massive star
\citep[e.g.][]{woosley93} and coelescence of two neutron
stars (NS-NS merger) or one neutron star and one black hole (NS-BH merger)
\citep[e.g.][]{paczynski86,eichler89}. 
A large angular momentum and a strong magnetic field are essential to launch
a jet \citep[e.g.][]{rezzolla11,etienne12}.
There are two types of plausible central engine:
one is a promptly formed black hole \citep[e.g.][]{popham99}, which accretes 
materials from the remnant with an extremely-high accretion rate 
($\sim (0.1-1) {\rm M_\odot~s^{-1}}$); the other is a strongly magnetized 
(with surface magnetic field $\sim 10^{15}$ G) neutron star which is spinning 
near the break-up limit (millisecond rotation period) \citep[e.g.][]{usov92}.
Our FRB model invokes this latter central engine. 

Even without direct evidence, a magnetar central engine is inferred indirectly
for some GRBs.
A shallow decay phase (or ``plateau'', Fig.1 lower panel) in the early X-ray 
afterglow of most long GRBs may require continuous energy injection into the 
blastwave \citep{zhang06},
which would be consistent with a spinning-down neutron star engine
\citep{dailu98b,zhangmeszaros01a}. An alternative explanation does not
invoke a long-lasting central engine, but invokes a stratification
of the ejecta Lorentz factor \citep{rees98}. The degeneracy 
between the two models was broken when
the so-called ``internal plateaus'' were discovered in the early X-ray 
afterglow lightcurves of some GRBs (Fig.1 upper panel) \citep{troja07,liang07b}. 
These are X-ray plateaus followed by an extremely steep decay, with decay
index steeper than -3, sometimes reaching -9. By contrast, most ``normal''
plateaus are followed by a decay with a decay index around -1, which is
well consistent with the external shock model of GRBs \citep[e.g.][for a 
recent review of the external shock model of GRBs]{gao13d}. The steepest
decay allowed in the external shock model is defined by high-latitude 
emission of a relativistic ejecta (e.g. when the blastwave enters a density
void), which has a decay slope $\alpha = -2 + \beta$ (convention
$F_\nu \propto t^{\alpha} \nu^{\beta}$), which usually cannot be smaller
than -3 \citep{kumar00}. The very steep decay following those internal
plateaus therefore demands an internal dissipation mechanism (rather than
the external shock emission) to account for the data (and hence, the plateaus
gain their name). This demands that the central engine lasts much longer
than the burst duration. The essentially constant X-ray luminosity during
the plateau requires a steady central engine output, and a spinning-down
magnetar naturally accounts for the data. Later a systematic analysis revealed 
more internal plateaus \citep{lyons10}. Surprisingly, such a signature was also 
found in a good fraction of short GRBs \citep{rowlinson10,rowlinson13}.

%Theoretically, how a magnetar engine may power a long GRB within the 
%massive star core collapse scenario has been studied extensively 
%\citep[e.g.][]{usov92,bucciantini09,metzger11}. In general, the 
%proto-neutron star needs to cool down first before a highly magnetized
%wind is launched. Within the NS-NS merger scenario, a supra-massive
%neutron star engine has been invoked to interpret delayed activities,
%such as flares \citep{dai06}, energy injection in the shallow decay 
%phase \citep{fanxu06}, and extended emission \citep{metzger08}. The
%intial short/hard spike of short GRBs is not straightforwardly expected
%in the magnetar engine scenario, but brief early accretion 
%\citep{metzger08} or viscous adjustment of neutron star from 
%differential rotation to uniform rotation \citep{fan13b} have been
%proposed. 

If one accepts that a millisecond magnetar is indeed operating in 
both long \citep{usov92,bucciantini09,metzger11} and short
\citep{dai06,fanxu06,metzger08,kiuchi12} GRBs\footnote{Within the magnetar
central engine model, a short GRB may be produced through a brief
accretion phase \citep{metzger08}, a brief differentially rotating
phase \citep{fan13c}, or a rapid phase transition phase
\citep[e.g.][]{dailu98b}.},
then the very steep decay at the end of internal
plateaus suggests that the emission stops abruptly. It is difficult
to turn off a rapidly spinning-down magnetar unless it collapses 
into a black hole. Simulations show that a rapidly spinning neutron
star can have a threshold mass (for collapsing into a black hole)
that is larger by 30\% - 70\% (depending on equation-of-state)
than the maximum mass of a non-rotating 
neutron star \citep{bauswein13}. As a result, in a large parameter
space, it is possible that a proto- millisecond magnetar born
with a baryon mass somewhat larger than the
maximum mass of a non-rotating neutron star may undergo rapid
spindown within $10^3 - 10^4$ s, 
and collapse into a black hole when it loses centrifugal support.
The collapsing time may be near the dipole spindown time scale $\tau \sim 2\times 10^3
~{\rm s}~ I_{45} B_{p,15}^{-2} P_{0,-3}^2 R_6^{-6}$, where $I$, $B_p$,
$P_0$ and $R$ are moment of inertia, surface magnetic field at the pole,
initial spin period at birth, and radius of the neutron star,
respectively, and the convention $Q_x = Q/10^x$ has been adopted
in cgs units. This can happen both in massive star core collapses
\citep{troja07} and NS-NS mergers \citep{zhang13,gao13a,yu13}.

As a supra-massive neutron star collapses into a black hole, magnetic
``hair'' has to be ejected based on the no-hair theorem of black hole.
The strong magnetic fields in the magnetar magnetosphere would reconnect
and get detached from the event horizon and expelled in a catastrophic manner.
The total energy in the magnetic field can be estimated as ($R_{\rm LC} \gg R$
is the light cylinder radius)
\begin{eqnarray}
E_{\rm B} & \simeq & \int_R^{R_{\rm LC}} (B_p^2/8\pi)^2 (r/R)^{-6} 4\pi r^2 dr \nonumber \\
& \simeq & (1/6) B_p^2 R^3 = 1.7\times 10^{47}~{\rm erg}~ B_{p,15}^2 R_6^3.
\label{EB}
\end{eqnarray} 
This is much larger than the observed energy of FRBs. Only a small amount
of this energy is adequate to power an FRB. 

The conversion of a small fraction of this energy to radio emission energy
invokes a poorly known coherent radio emission mechanism, such as
coherent curvature radiation through ``bunches''
\citep{RS75,falcke13} or ``maser''-like amplifications of plasma modes
\citep[e.g.][]{melrose09}. Here we assume that a certain coherent
mechanism can operate during the hair-ejection process, and the observed 
$\nu_{\rm obs}=(1.2-1.5)$ GHz radio wave can escape the emission 
region\footnote{We note that the condition for the observed frequency
to be above plasma frequency defined by the \cite{goldreich69} density 
is not required, since a force-free pulsar magnetosphere is charge-separated.
In fact, the 400 MHz ``core'' radio emission from pulsar polar cap region
\citep{rankin83} is below the plasma frequency defined 
by the Goldreich-Julian density but is observed.}. It can reach the observer 
if it is not absorbed by the GRB blastwave in front of the FRB emission region.
The comoving electron number density in the shocked ejecta region of the 
blastwave is $n' \simeq 1.8\times 10^5~{\rm cm}^{-3} L_{52} \Gamma_2^{-2}
r_{17}^{-2}$ ($L$ is the wind luminosity of the GRB, $\Gamma$ bulk Lorentz
factor, and $r$ the blastwave radius), which gives a comoving plasma frequency 
$\nu'_p = (n'e^2/\pi m_e)^{1/2} \simeq 3.8\times 10^6 ~{\rm Hz}~ L_{52}^{1/2}
\Gamma_2^{-1} r_{17}^{-1}$, much smaller than the radio wave
frequency in the comoving frame $\nu_{\rm obs}/\Gamma = (1.2-1.5) \times
10^7$ Hz. The density in the shocked circumburst medium region is even
lower. So the FRB emission can pass through the blastwave region and reach
Earth. Overall, the magnetic bubble associated with this FRB ejection, 
with total energy described by Eq.(\ref{EB}), would accelerate and convert 
the energy to the kinetic form. 
This energy is however small compared with the GRB energy,
so would not leave noticeable imprint in the GRB afterglow light curve.

{\em Since a sharp drop at the end of an ``internal X-ray plateau'' marks sudden 
cessation of the central engine, we suggest that the break time 
(or shortly after) is the epoch when an FRB is emitted} (Fig.1). 

\section{FRBs with and without GRBs}

The observed FRB event rate ($R_{\rm FRB} \sim 10^{-3}~{\rm gal}^{-1}
~{\rm yr}^{-1}$) is almost 3 orders of magnitude higher than long GRBs,
which is $R_{\rm GRB} \sim {\rm several}~ 10^{-6}~{\rm gal}^{-1} ~{\rm yr}^{-1}$ at
cosmological distances. So the scenario discussed above cannot account
for all FRBs. {\em Only a small fraction of FRBs could be associated with
GRBs.} Also since not all GRBs would have a supra-massive millisecond
magnetar as the central engine, {\em not all GRBs would be associated with FRBs}.

What is the nature of the majority of FRBs? 
We first consider the possibility that they are the same systems similar to 
GRB-associated FRBs but viewed at an off-jet angle. 
Even though these events may marginally account for the event rate \citep{frail01},
the FRBs (even if generated) are most likely not detectable.
For massive star (long) GRBs, it 
is believed that a more isotropic supernova should accompany the GRB,
which would screen any radio signal from the central engine. Compact
star (short) GRBs may be more transparent. An internal X-ray plateau due
to magnetar wind dissipation would have a near isotropic emission pattern,
which could be observed without a short GRB association \citep{zhang13}.
However, any FRB emission would still be absorbed by the ejecta
launched during the merger process, which has a mass at least
$10^{-4} M_\odot$ \citep[e.g.][]{freiburghaus99,rezzolla10,hotokezaka13}.
At about 1000 s (typical time at the end of internal plateaus)
\citep{rowlinson10,rowlinson13}, this ejecta has traveled a distance
$r \sim 0.2 c \times 1000~{\rm s}~ t_{3} = 6\times 10^{12} t_3$ cm. With a 
width $\Delta \sim 10^7$ cm, the mass density of the ejecta is
$\rho = M/(4\pi r^2 \Delta) \sim 4.4\times 10^{-4}~{\rm g~cm^{-3}}
M_{-3} t_3^{-2} \Delta_7^{-1}$. Assuming an average atomic number $Z$
for the ejecta, the electron number density is 
$n_e \sim 2.6 \times 10^{20} \chi Z^{-1} M_{-3} t_3^{-2} \Delta_7^{-1} ~{\rm cm}^{-3}$, 
where $\chi$ is the ionization fraction. The plasma frequency is $\nu_p
\sim 1.5\times 10^{14}~{\rm Hz}~ \chi^{1/2}Z^{-1/2} M_{-3}^{1/2} t_3^{-1} \Delta_7^{-1/2}$, 
which is $\gg \nu_{\rm obs}$
for a reasonable ionization fraction $\chi$. So the FRB emission would be 
blocked by the ejecta. 
Such an off-axis FRB may be still observable if the ejecta 
has a ``filling'' factor less than unity, so that the FRB emission can be visible
at certain solid angles. Nonetheless, this off-axis model can at most account
for a small fraction of FRBs not associated with GRBs.

This leaves the conclusion that 
most FRBs are implosions with a much longer 
delay (thousands to million of years, Falcke \& Rezzolla, 2013). The environment
is clean. The FRBs, once generated, can escape the source and reach Earth.
However, since the magnetic field strength 
is typically several times of $10^{12}$ G, the total energy of the 
explosion is smaller by 5-6 orders of magnitude than Eq.(\ref{EB}).
This energy would be converted to kinetic energy of an outflow and then
drives an ``afterglow'' of the FRB. Due to low baryon contamination, 
this outflow can reach a high Lorentz factor. However, due to its low
energetics, the afterglow would be too faint to be detectable.
According to the synchrotron external shock model of GRBs 
\citep[e.g.][]{meszarosrees97,sari98,gao13d}, the peak spectral
density of the afterglow is directly proportional to the energy of
the fireball, which can be scaled as 
$F_{\nu,max} = 1.1\times 10^{-6} ~{\rm \mu Jy}~ E_{42} n^{1/2}
\epsilon_{B.-2}^{1/2} D^{-2}_{28} [(1+z)/2]$. 
The deceleration time is short, $t_{\rm dec} = (3E/16\pi n m_p 
\Gamma_0^8 c^5)^{1/3} \simeq 0.1~{\rm s}~ E_{42}^{1/3} \Gamma_{0,2}^{-3/8}
[(1+z)/2]$.
The characteristic synchrotron frequency is $\nu_m = 4.3\times 10^{11}
~{\rm Hz}~ E_{42}^{1/2} \epsilon_{e,-1}^2 \epsilon_{B,-2}^{1/2} t_1^{-3/2}
[(z+1)/2]^{1/2}$.
Here $E$ is the total energy in the blastwave, $n$ is the ambient proton
number density, $\epsilon_e$ and $\epsilon_B$ are fractions of the shocked
energy that are distributed to electrons and magnetic fields, respectively,
$\Gamma_0$ is the initial Lorentz factor, $t$ is the observer time, and
$D$ is the luminosity distance.
Even for a magnetar with a total energy budget $E \sim 2\times 10^{47}$
erg (Eq.(\ref{EB})), the peak flux density can only reach $F_{\nu,max} \sim 
{\rm \mu Jy}$ at $z=0.5$. The deceleration time is $t_{\rm dec} \sim 3$ s, and 
the 1.3 GHz light curve reaches the peak (${\rm \mu Jy}$) at around
340 s after the FRB.

Since the X-ray internal plateau was detected in a good fraction of short GRBs,
and since NS-NS merger has been regarded as a top candidate to power
short GRBs, our picture suggests that some FRBs are also associated
with gravitational wave bursts (GWBs) due to NS-NS mergers. 
Our picture is different from Totani (2013), who suggested that all
FRBs are associated with NS-NS mergers. In his picture, an FRB is 
generated due to interaction of the magnetospheres of the two neutron stars, 
while in our picture, it happens 100s to 1000s seconds after the merger. 
As a result, the FRB signal in our picture would be blocked by the ejecta
launched during the merger in a large solid angle, so that the rate of 
FRB/GWB associations in our picture is much less than that of 
Totani (2013).

\section{Multi-wavelength/multi-messenger observational campaign
to unveil the nature of FRBs}

Within the framework delineated in this paper, 
one would consider the following strategies to
unveil the nature of FRBs. 

1. Since some GRBs may have generated an FRB $10^2 - 10^4$ s after the
GRB trigger, a prompt radio follow-up of GRBs would be essential to
verify or rule out our proposal. The fraction of long GRBs
that show an internal plateau is low, i.e. $\sim 3\%$, \citep{lv13b}.
So the chance of catching an internal plateau
would be low. However, an internal plateau is obervable only if the
external shock emission is relatively weak (Fig.1). Many ``normal'' 
plateaus could be also related to magnetars, since their decay slopes
and spectral indices are consistent with
being due to energy injection of a millisecond magnetar \citep{lv13b}. 
The end of plateau could be interpreted as the spindown
time scale of a magnetar, some of which may be related to implosion. 
These normal plateaus would outshine the internal plateaus if
the afterglow level is high.
One would then expect that a fraction of normal plateaus would also
be accompanied by FRBs at the end of the plateaus (or shortly after). 
The fraction of these
magnetar-candidate normal plateaus can be up to $\sim 60$\% of the 
long GRB population \citep{lv13b}. For short GRBs,
it seems that the chance of catching an internal plateau is much
higher \citep{rowlinson13}, although the end of plateau is earlier
(100s of seconds). In the past, radio follow up observations of GRBs
have been carried out much later, partially because of the technical
challenge for rapid slew but also partially because of the lack of
theoretical motivation (the predicted radio afterglow peaks days to 
weeks after a GRB trigger). The possible FRB/GRB connection proposed
here hopefully would give more impetus for prompt radio follow-up of GRBs.

2. Most FRBs are not supposed to be associated with GRBs. Many of these
will be discovered with future wide-field array searches 
\citep[e.g.][]{trott13}. 
Broad-band follow-up observations of these FRBs are encouraged. 
Possible detections may be made if a small fraction of these FRBs are 
associated with off-axis GRBs, which may be possible for NS-NS mergers
leaving behind broad-band (afterglow and mergernova) signals powered by 
a pre-collapsing magnetar \citep{zhang13,gao13a,yu13}. 
For most FRBs, the afterglow would be too weak to detect, unless they
are very nearby (peak flux ${\rm \mu Jy}$ 
at $z=0.5$ for a magnetar).

3. Nearby short GRBs with FRBs would be accompanied by GWBs, both before
the short GRB (in-spiral signal), between the short GRB and the FRB
(e.g. secular bar-mode instability of the supra-massive neutron star),
and shortly after the FRB (ring-down). Advanced LIGO/Virgo may be able
to detect these signals if the source is close enough. Some FRBs may be 
also associated with GWBs without a short GRB association, if the
ejecta filling factor of NS-NS mergers is not too large.

\section{Summary}

Along the line of Falcke \& Rezzolla (2013) who proposed that FRBs can
be produced when a supra-massive neutron star loses centrifugal support
and collapses into a black hole, here we suggest that a small fraction
of such implosions can happen shortly ($10^2 - 10^4$ s) after the 
formation of a supra-massive neutron star, which could produce an FRB
around an X-ray break time following some GRBs, both long and short (Fig.1).
Not all GRBs could
make FRBs, but a good fraction could do. We therefore suggest a prompt radio
follow-up observation for GRBs, and suggest that observations as early
as 100 s after GRB triggers would be useful. If observations cover
the period when the X-ray plateau (both ``internal'' and ``normal'',
Fig.1) is observed, a detection or non-detection of an FRB at the end
of the plateau (or shortly after) would greatly constrain the nature of FRBs. 
Most FRBs are not supposed to be associated with GRBs. In any case, some
faint signals are predicted, and multi-wavelength follow-up observations 
of FRBs may lead to detection of these signals under
optimistic circumstances.

After posting the first version of this paper to arXiv, I was informed
(2013, A. van der Horst, private communication) that the suggested FRB/GRB
association may have been detected. In a conservative paper where the
authors reported an upper limit of early radio afterglow flux, \cite{bannister12}
reported two FRB-like events following two long GRBs, at an epoch
close to what is predicted in this paper. Further dedicated observations
are needed to unveil the rich GRB/FRB association phenomenology.

%% In this section, we use  the \subsection command to set off
%% a subsection.  \footnote is used to insert a footnote to the text.

%% Observe the use of the LaTeX \label
%% command after the \subsection to give a symbolic KEY to the
%% subsection for cross-referencing in a \ref command.
%% You can use LaTeX's \ref and \label commands to keep track of
%% cross-references to sections, equations, tables, and figures.
%% That way, if you change the order of any elements, LaTeX will
%% automatically renumber them.

%% This section also includes several of the displayed math environments
%% mentioned in the Author Guide.

\acknowledgments
%I thank a Cheung Kong scholarship of China, the hospitality of the 
%KIAA and Department of Astronomy of Peking University, and the 
%sabbatical committee of the UNLV faculty senate, to provide me an 
%ideal working environment to conduct research efficiently.  
I thank Matthew Bailes, Alexander van der Horst,
Peter M\'esz\'aros, Kunihito Ioka, Kazumi 
Kashiyama, Dong Lai, and Feng Yuan for helpful discussion and an
anonymous referee for important remarks.
% for discussion on FRB observations, and 
%Dong Lai and Feng Yuan for discussions on possible physical mechanisms 
%to produce FRBs. 
%stimulative discussion with Xue-Feng Wu, He Gao, Zi-Gao Dai,
%and Yi-Zhong Fan, and helpful comments from Kunihito Ioka, Elenora
%Troja and an anonymous referee. 
This work is partially supported by NASA NNX10AD48G and NNX11AQ08G.

\begin{figure}
\epsscale{.80}
\plotone{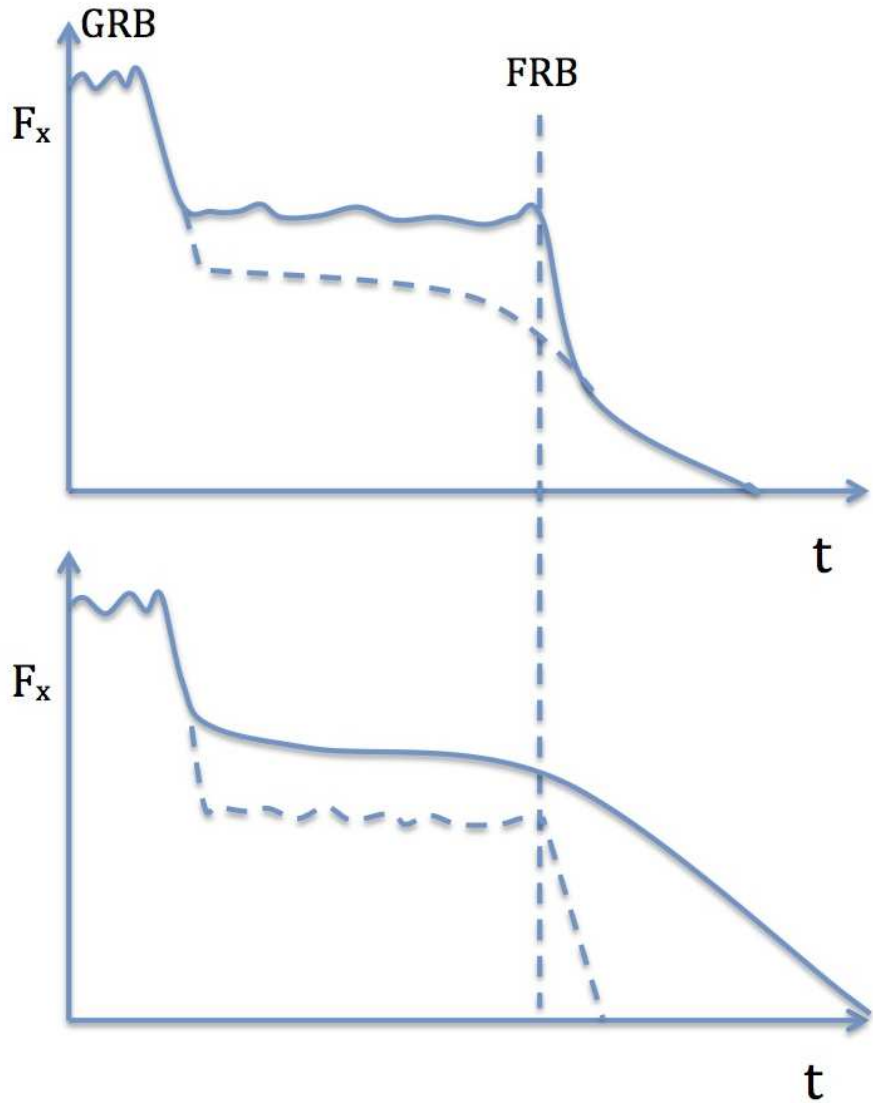}
\caption{A cartoon picture of FRB/GRB connection. Two indicative 
X-ray afterglow light curves are presented. The upper panel shows
an ``internal plateau'', with a near-steady plateau emission followed
by a very steep decay. 
%The emission has to be of an internal origin,
%likely due to magnetic energy dissipation of a spinning down magnetar.
The end of the plateau (or shortly after) may signal
collapse of the supra-massive magnetar
into a black hole (dashed vertical line). We suggest this epoch as
the emission epoch of an FRB. The external shock emission of these
lightcurves (dashed curve) is buried below the internal plateau.
The lower panel shows a ``normal'' plateau, which is dominated by
the external shock emission. The end of plateau may also coincide with
the end of magnetar energy injection, a fraction of which could be 
also due to magnetar implosion. The internal dissipation emission  
of these cases (dashed curve) is outshone by the external shock 
emission. The break time (or shortly after) of some of these normal 
plateaus could also coincide with an FRB.
 \label{fig1}}
\end{figure}

%% Use the figure environment and \plotone or \plottwo to include
%% figures and captions in your electronic submission.
%% To embed the sample graphics in
%% the file, uncomment the \plotone, \plottwo, and
%% \includegraphics commands
%%
%% If you need a layout that cannot be achieved with \plotone or
%% \plottwo, you can invoke the graphicx package directly with the
%% \includegraphics command or use \plotfiddle. For more information,
%% please see the tutorial on "Using Electronic Art with AASTeX" in the
%% documentation section at the AASTeX Web site,
%% http://www.journals.uchicago.edu/AAS/AASTeX.
%%
%% The examples below also include sample markup for submission of
%% supplemental electronic materials. As always, be sure to check
%% the instructions to authors for the journal you are submitting to
%% for specific submissions guidelines as they vary from
%% journal to journal.

%% This example uses \plotone to include an EPS file scaled to
%% 80% of its natural size with \epsscale. Its caption
%% has been written to indicate that additional figure parts will be
%% available in the electronic journal.

\end{document}